\documentclass[iop,twocolumn]{emulateapj}

\newcounter{species}

\usepackage{color}
\usepackage{natbib}
\usepackage{graphicx}
\usepackage{rotating}

\shorttitle{New Red Jewels in Coma Berenices}
\shortauthors{Ryan C. Terrien et al.}
\begin{document}
\title{New Red Jewels in Coma Berenices}

\author{Ryan C. Terrien\altaffilmark{1,2,3},
Suvrath Mahadevan\altaffilmark{1,2,3}, 
Rohit Deshpande\altaffilmark{1,2}, 
Chad F. Bender\altaffilmark{1,2},
Phillip A. Cargile\altaffilmark{4},
Frederick R. Hearty\altaffilmark{5,1},
Michiel Cottaar\altaffilmark{6},
Carlos Allende Prieto\altaffilmark{7,8},
Scott W. Fleming\altaffilmark{9},
Peter M. Frinchaboy\altaffilmark{10},
Kelly M. Jackson\altaffilmark{11,10},
Jennifer A. Johnson\altaffilmark{12,13},
Steven R. Majewski\altaffilmark{5},
David L. Nidever\altaffilmark{14,5},
Joshua Pepper\altaffilmark{15,4},
Joseph E. Rodriguez\altaffilmark{4},
Donald P. Schneider\altaffilmark{1},
Robert J. Siverd\altaffilmark{4},
Keivan G. Stassun\altaffilmark{4,16},
Benjamin A. Weaver\altaffilmark{17},
\& John C. Wilson\altaffilmark{5}
}
\slugcomment{ApJ Accepted, Jan 2014}

\email{rct151@psu.edu}
\altaffiltext{1}{Department of Astronomy and Astrophysics, The Pennsylvania State University, 525 Davey Laboratory, University Park, PA 16802, USA.}
\altaffiltext{2}{Center for Exoplanets and Habitable Worlds, The Pennsylvania State University, University Park, PA 16802, USA.}
\altaffiltext{3}{The Penn State Astrobiology Research Center, The Pennsylvania State University, University Park, PA 16802, USA.}
\altaffiltext{4}{Department of Physics and Astronomy, Vanderbilt University, VU Station 1807, Nashville, TN 37235, USA.}
\altaffiltext{5}{Department of Astronomy, University of Virginia, P.O. Box 400325, Charlottesville, VA 22904-4325, USA.}
\altaffiltext{6}{Institute for Astronomy, ETH Z\"{u}rich, Wolfgang-Pauli-Strasse 27, CH-8093, Zurich, Switzerland.}
\altaffiltext{7}{Instituto de Astrof\'{i}sica de Canarias (IAC), C/V\'{i}a L\'{a}ctea, s/n, E-38200, La Laguna, Tenerife, Spain}
\altaffiltext{8}{Departamento de Astrof\'{i}sica, Universidad de La Laguna, E-38206, La Laguna, Tenerife, Spain.}
\altaffiltext{9}{Space Telescope Science Institute, 3700 San Martin Dr, Baltimore MD 21211, USA}
\altaffiltext{10}{Department of Physics \& Astronomy, Texas Christian University,
TCU Box 298840, Fort Worth, TX 76129, USA.}
\altaffiltext{11}{Department of Physics, University of Texas--Dallas, Dallas, TX 75080, USA.}
\altaffiltext{12}{Department of Astronomy, Ohio State University, 140 West 18th Avenue, Columbus, OH 43210, USA.}
\altaffiltext{13}{Center for Cosmology and Astro-Particle Physics, Ohio State University, Columbus, OH 43210, USA.}
\altaffiltext{14}{Department of Astronomy, University of Michigan, Ann Arbor, MI, 48104, USA.}
\altaffiltext{15}{Department of Physics, Lehigh University, Bethlehem, PA 18015, USA.}
\altaffiltext{16}{Department of Physics, Fisk University, 1000 17th Ave. North, Nashville, TN, 37208, USA.}
\altaffiltext{17}{Center for Cosmology and Particle Physics, New York University, New York, NY 10003, USA.}

\begin{abstract}
We have used Sloan Digital Sky Survey-III (SDSS-III) Apache Point Observatory Galactic Evolution Experiment (APOGEE) radial velocity observations in the near-infrared $H$-band to explore the membership of the nearby ($86.7 \pm 0.9$~pc) open cluster Coma Berenices (Melotte~111), concentrating on the poorly-populated low-mass end of the main sequence. Using SDSS-III APOGEE radial velocity measurements, we confirm the membership of eight K/M dwarf members, providing the first confirmed low-mass members of the Coma Berenices cluster. Using $R\sim2000$ spectra from IRTF-SpeX, we confirm the independently luminosity classes of these targets, and find their metallicities to be consistent with the known solar mean metallicity of Coma Berenices and of M dwarfs in the Solar neighborhood. In addition, the APOGEE spectra have enabled measurement of $v\sin i$ for each target and detection for the first time of the low-mass secondary components of the known binary systems Melotte~111~102 and Melotte 111 120, as well as identification of the previously unknown binary system 2MASS J12214070+2707510. Finally, we use Kilodegree Extremely Little Telescope (KELT) photometry to measure photometric variability and rotation periods for a subset of the Coma Berenices members.

\end{abstract}


\section{Introduction and Background}
Star clusters are invaluable astrophysical laboratories and nearby open clusters in particular provide a wealth of information about stellar evolution and dynamics. Constraints on cluster compositions and ages can be leveraged to inform understanding of the formation and evolution of stellar binary systems. The measurement of the frequencies and mass ratios of binary systems in clusters, which have a common age and history, can clarify how different dynamical processes contribute to their development \citep[][and references therein]{Duchene:2013il}. A similar process can be brought to bear on questions about the formation and evolution of planetary systems. Currently, there are seven cluster members known to host planets \citep{Sato:2007fy,Lovis:2007cy,Quinn:2012co,2013Natur.499...55M,Quinn:2013ur}, and there are indications that small planets may be as frequent within clusters as in the field \citep{2013Natur.499...55M}. 

Nearby clusters have also played a large role in refining the distance scale, for example, through comparisons between parallax measurements and stellar model-based isochrone fitting. The most famous example of this is the Pleiades, for which a significant discrepancy persists between \textit{Hipparcos} parallax measurements on one hand \citep{vanLeeuwen:2009db}, and model-based distance measurements from isochrone fitting \citep[e.g.][]{Pinsonneault:1998iq,Percival:2005hf} and eclipsing binaries \citep[e.g.][]{Southworth:2005em} on the other. As new parallax measurements are obtained and stellar models are developed, nearby clusters will remain crucial anchors for the distance ladder and the understanding of stellar evolution and structure. 

Confirmed new cluster members of nearby clusters are also prime targets for high contrast imaging searches targeted at detecting massive exoplanets. Knowledge of the age of the system is essential to translate the luminosity of the detected planet candidates into planetary mass. Surveying young stellar systems with known ages (i.e.\ cluster members) can help avoid ambiguities in the interpretation of companion mass \citep[see {e.g.}][]{Carson:2013fw,Hinkley:2013vf}.

The Coma Berenices cluster (Coma Ber, Melotte 111, $\alpha_{\mathrm{J2000}} = 12^{\mathrm{h}}23^{\mathrm{m}}$, $\delta_{\mathrm{J2000}} = +26^{\circ} 00^{\prime}$) is among the nearest open clusters. At $86.7 \pm 0.9$~pc \citep{vanLeeuwen:2009db}, Coma Ber is farther than the Hyades (45~pc) but closer than Praesepe (180~pc); its age of $\sim500$ Myr is similar to that of the Hyades ($\tau_{\mathrm{Hyades}}\sim600$ Myr) and Praesepe ($\tau_{\mathrm{Praesepe}}\sim500$ Myr) clusters \citep{Kharchenko:2005dw}. However, with $\lesssim 100$ known or suspected members, Coma Ber is substantially sparser than these other two clusters, which each contain several hundred members. 

Historically, discovery and verification of members of Coma Ber has been challenging due to its sparsity and its low proper motion (PM) (14.9~mas~yr$^{-1}$). \citet{Trumpler:1938vy} identified 37 of the highest-mass Coma Ber members, but subsequent work has only revealed a small number of additional members \citep{Argue:1969wh,Bounatiro:1993ws,Odenkirchen:1998gp}. The most recent photometric and PM studies \citep{Casewell:2006ga,Kraus:2007cx} present many high-probability candidate members, but do not confirm membership with radial velocities (RVs). 

The low-mass end of the main sequence of Coma Ber is currently poorly constrained, since the targets are optically faint. The list of confirmed members includes only a few early K dwarfs \citep{Mermilliod:2008dx} and no M dwarfs, although \citet{Kraus:2007cx} presented 76 M dwarfs and 43 K dwarfs with high membership probability based on PMs and photometry. These low-mass members are of particular importance to studies of cluster dynamics, as low-mass stars are expected to be the first to evaporate from a cluster. In Coma Ber, the uncertainty whether the low mass members have evaporated or were simply undiscovered has motivated several PM and RV searches \citep{Casewell:2006ga,Mermilliod:2008dx,Melnikov:2012ks}. A search for low-mass members identified through X-ray emission also yielded no confirmed members \citep{Randich:1996tc,GarciaLopez:2000uy}.

In this study we use the $R\sim22,500$, multiplexed, $H$-band ($1.5$--$1.8\mu$m) Apache Point Observatory Galactic Evolution Experiment (APOGEE) spectrograph \citep{2008AN....329.1018A,2010SPIE.7735E..46W}, which is a part of the Sloan Digital Sky Survey-III \citep[SDSS-III;][]{2011AJ....142...72E} on the 2.5~m SDSS telescope \citep{Gunn:2006fu}, to provide RV confirmation of eight low-mass members of Coma Ber. These members include six previously-suspected M dwarf members, one new K/M dwarf member, and one new K/M dwarf binary member. We also derive double-lined spectroscopic binary orbits for two previously known binary members and the new K/M dwarf binary member. These M dwarfs are the first low-mass stars to be verified as members of the Coma Ber cluster with RVs. In Section \ref{secobs} we describe the APOGEE observations, in Section \ref{sectargsel} we describe the target selection, in Section \ref{secchar} we describe further characterization we performed, and in Section \ref{secres} we discuss the characteristics and significance of the new M dwarf members of Coma Ber.

\section{SDSS-III APOGEE Observations}
\label{secobs}
The spatial region of the Coma Ber cluster was specifically targeted by APOGEE as part of its main Galactic survey. Although a single APOGEE pointing (field of view diameter~=~3$^{\circ}$) cannot encompass the full $7.5^{\circ}$ diameter of Coma Ber, it does cover a significant fraction of the cluster. The SDSS-III Data Release 10 (DR10) dataset \citep{2013arXiv1307.7735A}\footnote{http://www.sdss3.org/dr10/}, which presents the first year of APOGEE survey data, contains 385 targets in the Coma Ber field (APOGEE field name 221+84, plate IDs 5621, 5622) \citep{Zasowski:2013ea}. This list of targets includes those presented by \citet{Kraus:2007cx} and those presented by \citet{Casewell:2006ga}, and the selection process is detailed in P. M. Frinchaboy et al.\ (in preparation) and \citet{Zasowski:2013ea}. The number of visits for each target in the Coma Ber field ranges from one to six, with the majority having three, and these targets were observed between January and March 2012.

The APOGEE DR10 dataset is a product of the APOGEE extraction pipeline (D.~L. Nidever et~al.\ in preparation) and the APOGEE Stellar Parameters and Chemical Abundances Pipeline \citep[ASPCAP;][M. Shetrone et~al., in preparation; A.~E. Gar\'{i}ca P\'{e}rez et~al., in preparation]{Meszaros:2013va}. The extraction pipeline provides spectra that are wavelength calibrated, telluric-corrected, and masked of strong sky lines. The spectra span the range 1.514--$1.696\mu$m, with two spectral gaps corresponding to the physical gaps between the detector arrays. The combined spectra for all targets in this paper have a signal-to-noise ratio (SNR)~$>100$ per half-resolution element.

For each target star, the APOGEE pipeline determines an array of stellar parameters, including T$_{\mathrm{eff}}$, $\log(g)$, and a variety of preliminary elemental abundance values. These parameters are based on a global fit with a wide range of stellar models \citep{Meszaros:2012fu}. The APOGEE pipeline also provides barycentric RVs based on cross-correlations with a large variety of template spectra. This iterative process and the $\chi^2$ template selection are described in detail in D.~L. Nidever et~al.~(in preparation).

\section{Target Selection: Proper Motions, Radial Velocities, and Photometry}
\label{sectargsel}
\begin{figure*}
\begin{center}
\includegraphics[scale=0.5]{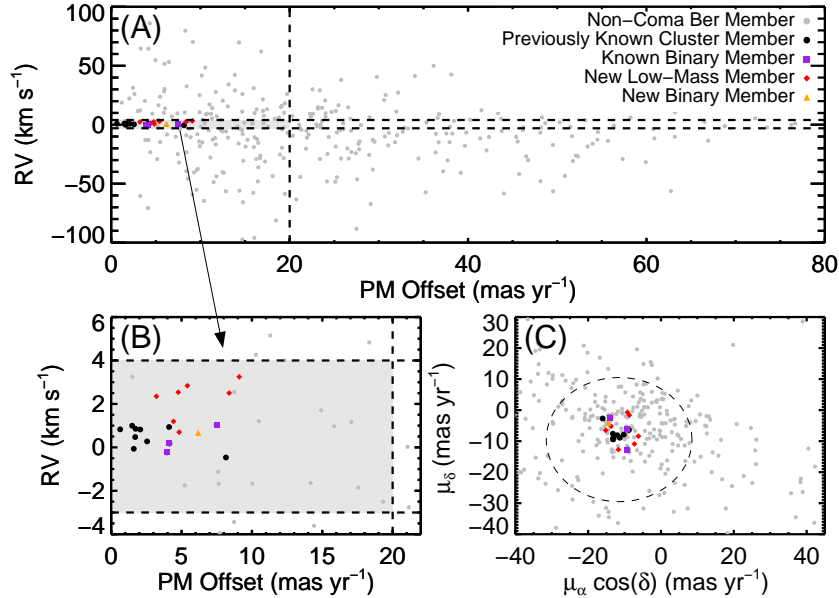}
\caption{(A) and (B) The distribution of APOGEE targets in PM and RV. The PM offset is the total offset from the cluster mean PM of $(\mu_\alpha \cos \delta, \mu_\delta) = (-11.5,-9.5)$~mas~yr$^{-1}$. The cut at 20~mas~yr$^{-1}$ is shown as a vertical dotted line. The RV cuts of $-3<\mathrm{RV}<+4$~km~s$^{-1}$ are shown as horizontal dotted lines. The RV values are the average of the RVs measured over all APOGEE epochs, except for the binary systems, for which the RV values are the known systemic ($\gamma$) velocities. (C) The distribution of the lowest proper motion APOGEE targets in proper motion space, using proper motion values from the PPMXL catalog \citep{Roeser:2010cr}. The cut at 20~mas~yr$^{-1}$ from the cluster mean proper motion $(\mu_\alpha \cos \delta, \mu_\delta) = (-11.5,-9.5)$~mas~yr$^{-1}$ is shown as a dotted circle. In all three frames, known cluster members are indicated, binary cluster members \citep{Mermilliod:2008dx} are highlighted, and our new targets are shown. Rejected candidates are addressed in the text and in Figure \ref{figphot}.
\label{figpmrv}}
\end{center}
\end{figure*}

The mean PM of Coma Ber is $(\mu_\alpha \cos \delta, \mu_\delta) = (-11.5,-9.5)$~mas~yr$^{-1}$ \citep{Kraus:2007cx}, and is not sufficiently different from the background population to provide a strong indicator of cluster membership by itself. However, PM remains a useful constraint for eliminating Solar neighborhood field stars and we follow \citet{Kraus:2007cx} in selecting targets with PMs within 20~mas~yr$^{-1}$ of the cluster mean. We drew PMs for the Coma Ber targets by cross-referencing with both the Position and Proper Motion Extended-L (PPMXL) catalog \citep{Roeser:2010cr} and the USNO CCD Astrograph Catalog 4 (UCAC4) \citep{Zacharias:2013cf}. Of the 385 targets in the APOGEE Coma Ber field, 177 satisfy the PM requirement above according to both the PPMXL and UCAC4 catalogs. The distribution of PMs is shown in Figure \ref{figpmrv}.

\begin{figure}
\begin{center}
\hspace{-20pt}\includegraphics[scale=0.4]{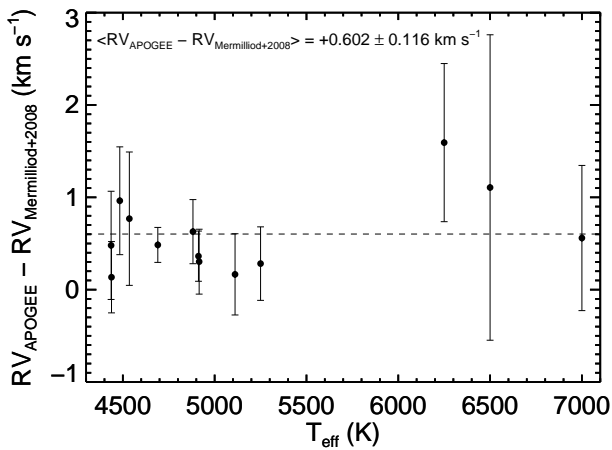} %fig 2
\caption{The difference between APOGEE and CORAVEL \citep{Mermilliod:2008dx} RV measurements for the set of 13 FGK stars observed with both instruments, organized by T$_{\mathrm{eff}}$ from ASCAP, or the best-fit RV template if the \texttt{STAR\_BAD} ASPCAP flag is set. Known binary stars are excluded. The APOGEE RVs are consistently higher than the CORAVEL RVs, with a mean offset of $0.602 \pm 0.116$~km~s$^{-1}$.  
\label{rvcomp}}
\end{center}
\end{figure}

The APOGEE reduction pipeline provides barycentric RVs, derived by cross correlation against spectral templates. For the lowest-mass dwarfs, the BT-Settl \citep{Allard:2012fp} model spectra are used as templates. For single stars with good SNR, D.~L. Nidever {et~al.} (in preparation) compare APOGEE RV measurements to literature measurements for globular cluster targets with accurate RVs to determine that the APOGEE RV accuracy is $\sim0.26\pm0.2$~km~s$^{-1}$. For the Coma Ber targets, we also compare the APOGEE RVs to those measured with the CORAVEL spectrograph \citep{Mermilliod:2008dx}. A total of 13 targets were observed with both APOGEE (DR10) and CORAVEL, excluding known binaries, and we find the mean velocity offset is $\langle \mathrm{RV}_{\mathrm{APOGEE}} - \mathrm{RV}_{\mathrm{Mermilliod+2008}}\rangle = -0.602$~km~s$^{-1}$, with an error in the mean of 0.116~km~s$^{-1}$ ($\sigma = 0.417$~km~s$^{-1}$). The RV offsets for these targets are shown in Figure \ref{rvcomp}. \citet{Mermilliod:2008dx} report RVs for Coma Ber in the ELODIE RV system defined by \citet{Udry:1999wc}, which has been compared to the Keck/Lick planet search RV system by \citet{Nidever:2002fc}.

Ongoing work by another APOGEE ancillary science team indicates the possibility of a $\sim~2$~km~s$^{-1}$ systematic redshift for the coolest dwarfs (T$_{\mathrm{eff}}< 3500$~K) (Cottaar,~M.\ and Covey,~K. 2013, private communication) in RVs derived from the BT Settl model spectra, hints of which are also seen in RVs derived with BT Settl templates from spectra observed with NIRSPEC on Keck \citep{2005AJ....129..402B}. This apparent shift is not present for stars with T$_{\mathrm{eff}}>3500$~K (e.g.~in Figure~\ref{rvcomp}) and is present across the $H$-band, indicating a possible systematic effect in the coolest BT Settl models. This apparent shift is relevant to the new Coma Ber members presented here, and motivates a careful consideration of the upper RV limit taken to confirm cluster membership for the coolest stars. To ensure that the RVs provided here are traceable back to the SDSS-III DR10 public release, we do not correct the measured RVs, but account for this apparent shift in our selection criteria described below.

For each target in the Coma Ber field, we examine the average RV over all epochs as reported in the APOGEE catalog. To identify possible binary targets for which these average RVs might be incorrect, we examine the $\sigma_{\mathrm{RV}}$ over all epochs and the visual shape of the cross-correlation function measured by the APOGEE pipeline. Targets with asymmetric cross-correlation functions or $\sigma_{\mathrm{RV}} \gtrsim 0.3-0.4$~km~s$^{-1}$ are flagged as potential binary targets.

\citet{Mermilliod:2008dx} find a mean barycentric RV for Coma Ber of $\langle V_{r} \rangle = 0.01 \pm 0.08$~km~s$^{-1}$, and elect to use a membership criterion of $-2 < \mathrm{RV} < 2$~km~s$^{-1}$ for their CORAVEL measurements. Given the lack of extensive heritage of the zero-points from APOGEE velocities spanning all spectral types and the offset from the ELODIE RV system, we impose a membership criterion of at least $-3<\mathrm{RV}<3$~km~s$^{-1}$, to ensure that we include most bona fide cluster members. To account for the possible RV shift for stars with T$_{\mathrm{eff}}<3500$~K detailed above, we conservatively increase the upper limit by 1~km~s$^{-1}$, resulting in a final RV membership criterion of $-3<\mathrm{RV}<4$~km~s$^{-1}$. There are 59 targets in the entire dataset which satisfy this RV condition; 28 of these also satisfy the PM criteria described above. These targets are shown in Figure \ref{figpmrv}.

Of the 28 stars that satisfy the RV+PM criteria, 16 do not have the \texttt{STAR\_BAD} ASCAP flag\footnote{APOGEE flags are detailed at http://www.sdss3.org/dr10/} set and therefore have acceptable S/R, model fit, and parameters that are within the ASPCAP grids (particularly T$_{\mathrm{eff}}>3500$K). Five of these 16 stars have 4500K~$<$~T$_{\mathrm{eff}}<5000$~K and $2.6<\log(g)<3.5$. Since these are likely giants, we do not include them in our list of candidate Coma Ber members. Four of these five stars are also excluded from cluster membership by the photometric criteria below. This leaves 23 candidate stars that satisfy the RV+PM criteria and do not have known low gravity.
 
Three known single-lined spectroscopic binary (SB1) members of Coma Ber were observed by APOGEE: Tr 120 (2MASS J12260547+2644385), Tr 102 (2MASS J12234182+2636054), and Tr 97 (2MASS J12230840+2551049) \citep{Mermilliod:2008dx}. The pipeline RVs easily identified each of these as highly RV variable targets even with only three epochs each. The $\sigma_{\mathrm{RV}}$ values for these targets are 7.22~km~s$^{-1}$, 4.21~km~s$^{-1}$, and 30.89~km~s$^{-1}$ respectively; these values differ significantly from the usual APOGEE scatter of 0.1 - 0.5~km~s$^{-1}$. Since these objects are spectroscopic binaries, the average RVs reported by the APOGEE pipeline are inaccurate, but we include them in our Coma Ber list because their systemic ($\gamma$) velocities have previously been measured and confirm their membership in Coma Ber \citep{Mermilliod:2008dx}. Thus we find 26 RV + PM + $\log(g)$ candidate members.

\begin{figure}
\begin{center}
\hspace{-7mm}\includegraphics[scale=0.40]{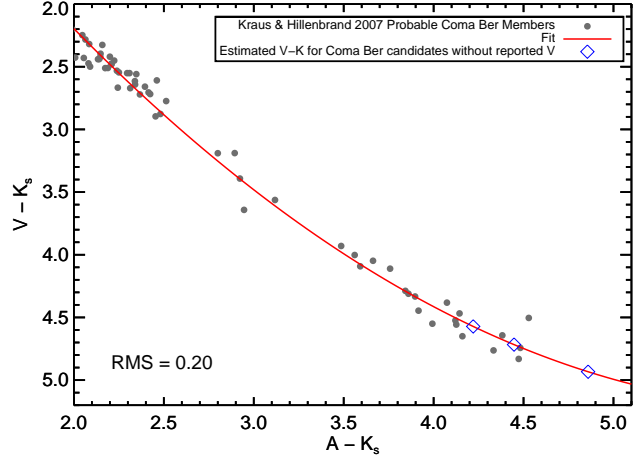} % fig 3
\caption{The tight relation between $V-K_{s}$ and $A-K_{s}$ for high-probability Coma Ber members reported by \citet{Kraus:2007cx}, where $A$ is the aperture magnitude from UCAC4. This relation is used to estimate $V-K_{s}$ for three stars without $V$ magnitudes in UCAC4. One target (2MASS J12231356+2602185, target ID 0) requires an extrapolation of the fit.}\label{figakvk}
\end{center}
\end{figure}

The final criterion we applied to identify members of Coma Ber was photometric. We compared the positions of the targets to Dartmouth Stellar Evolution Program \citep[DSEP,][]{2008ApJS..178...89D} and Yale-Yonsei \citep[YY,][]{Spada:2013cs} isochrones (T~=~500~Myr and d~=~86.7~pc) in $(V-K_s, J)$, $(V-K_s, V)$, $J-H, J$, and $J-K_s, J$ color-magnitude diagrams (CMDs). The temperature sensitivity of $V-K_s$ enables discrimination of cluster members in both color and magnitude in the $V-K_s$ CMDs, and while the isochrones are nearly vertical in the infrared CMDs, they demonstrate that the infrared photometry is at least consistent with the $V-K_s$ CMDs. The infrared $J$, H, and $K_s$ magnitudes are drawn from the 2MASS \citep{Skrutskie:2006hl} point source catalog and the $V$ magnitudes are from the AAVSO Photometric All Sky Survey and compiled in the UCAC4 catalog. Three of the low-mass targets (target IDs 0, 3, and 4 below) did not have $V$ magnitudes reported in UCAC4, so for $V$ and $V-K_s$ we instead use an estimated $V$ based on a third-order polynomial fit to the relationship between $V-K_s$ and $A-K$ for the stars presented as likely Coma Ber members in \citet{Kraus:2007cx}, where $A$ is the aperture magnitude reported in UCAC4. This fit and the three estimated $V-K_s$ colors are shown in Figure \ref{figakvk}. For the following photometric analysis, we estimate an error in $V-K_s$ for these targets of 0.20.

\begin{figure*}
\begin{center}
\includegraphics[scale=0.6]{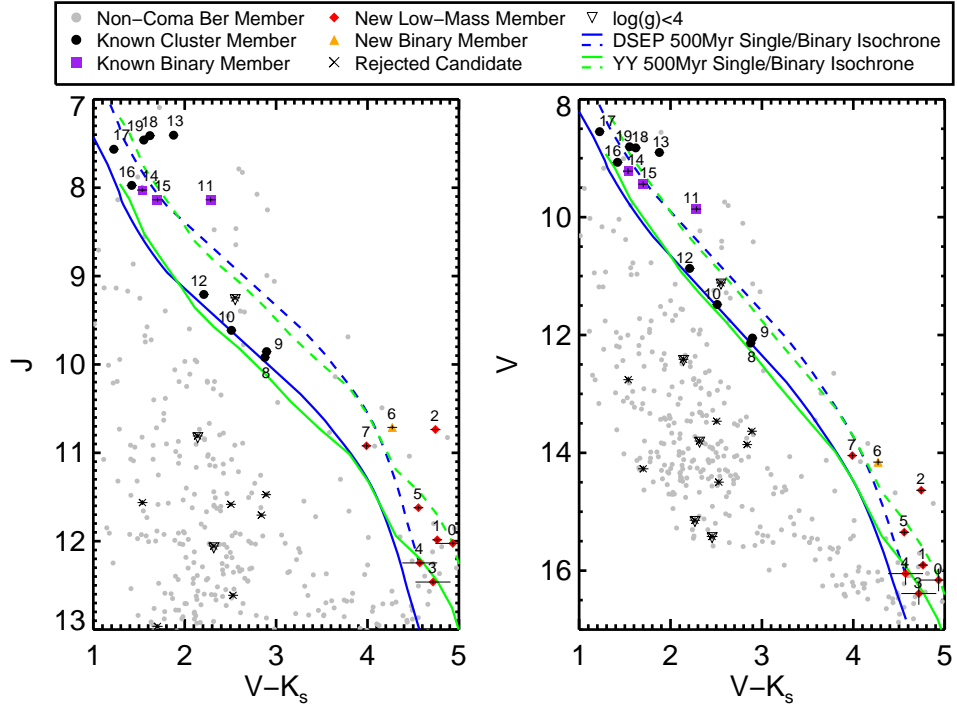} % fig 4
\caption{The $V-K_s, J$ and $V-K_s, V$ color-magnitude diagrams for the Coma Ber targets, along with 500~Myr DSEP and YY isochrones at a distance of 86.7 pc \citep{2008ApJS..178...89D,Spada:2013cs}. Targets are labeled by index in Table \ref{restable}, and previously-confirmed single and binary members are highlighted. The 11 targets that satisfy our PM and RV constraints but are not photometrically consistent with cluster membership or have $\log (g)<4$ are also shown.}\label{figphot}
\end{center}
\end{figure*}

\begin{figure*}
\begin{center}
\includegraphics[scale=0.6]{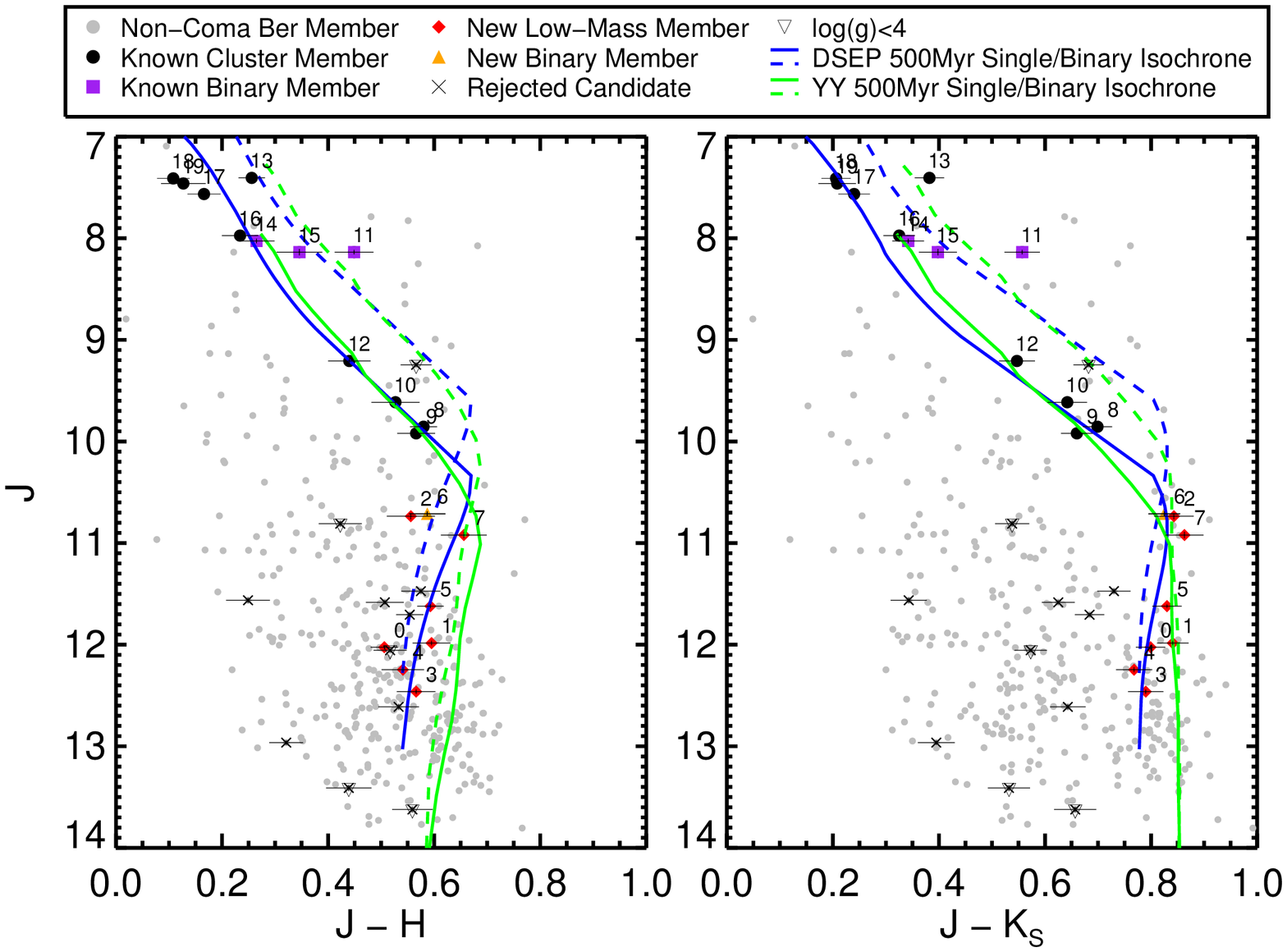} % fig 5
\caption{The $J-H, J$ and $J-K_s, J$ color-magnitude diagrams for the Coma Ber targets, along with 500~Myr DSEP and YY isochrones at a distance of 86.7 pc \citep{2008ApJS..178...89D,Spada:2013cs}. Targets are labeled by index in Table \ref{restable}, and previously-confirmed single and binary members are highlighted. The 11 targets that satisfy our PM and RV constraints but are not photometrically consistent with cluster membership or have $\log (g)<4$ are also shown.}\label{figphot2}
\end{center}
\end{figure*}

The Coma Ber CMDs, presented in Figures \ref{figphot} and \ref{figphot2}, show that 20 of the 26 RV + PM + $\log (g)$ selected targets have photometry consistent with the DSEP or YY single or binary isochrones and are therefore consistent with Coma Ber membership. There is a substantial spread in both the measured and model photometry for the reddest stars, but the Coma Ber low-mass stars fall within the wide range predicted by the models. The remaining six stars we reject as candidate members because they are much fainter than predicted by the isochrones. The most highly discrepant Coma Ber members include one known spectroscopic binary (Tr 120, target ID 11), and one target (Tr 92, target ID 13) that is listed as a Coma Ber member in \citet{Mermilliod:2008dx} but not in \citet{Kraus:2007cx}. The APOGEE observations of Tr 92 show a slightly higher than average RV scatter ($\sigma_{\mathrm{RV}}=1.15$~km~s$^{-1}$ and projected rotational velocity ($v\sin i = 25.50 \pm 5.20$~km~s$^{-1}$, derived in Section \ref{secvsini}), indicating possible multiplicity. 

\begin{figure*}
\begin{center}
\includegraphics[scale=0.6]{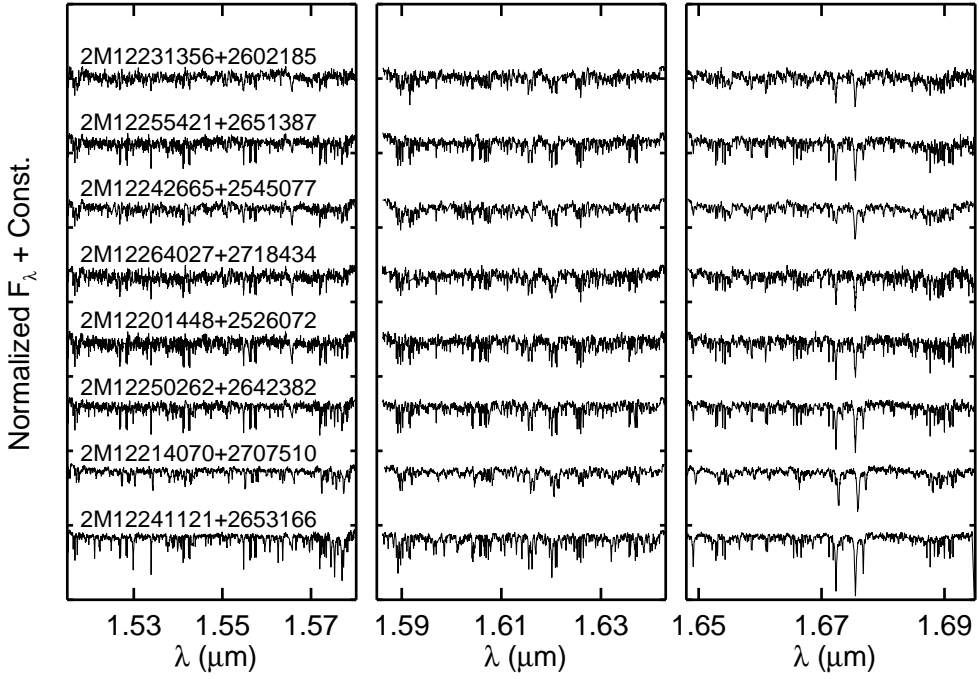} % fig 6
\caption{The combined normalized APOGEE spectra for each of the new low-mass stars in Coma Ber, including the new K/M binary. The left, middle, and right panels correspond respectively to the blue, green, and red detector chips of APOGEE. The spectra are arranged from bottom to top in order of decreasing effective temperature. The IDs are 2MASS IDs.}\label{figap}
\end{center}
\end{figure*}

The DSEP and YY isochrones are consistent with the known cluster members and have been shown in other clusters to provide reasonable fits to the lower main sequence \citep{2008ApJS..178...89D,Spada:2013cs}. However, they tend to predict different luminosities for a given color in the lowest mass stars. The seven new low-mass members and the new K/M dwarf binary presented here fall within the wide range of luminosities predicted by the DSEP and YY isochrones, and are well-distinguished from the rejected candidate members. The APOGEE spectra for these targets are shown in Figure \ref{figap}. The summary of information for all cluster members is shown in Table \ref{restable}.

%\begin{landscape}
%\begin{deluxetable*}{@{}p{.4cm}p{2.3cm}p{.6cm}p{.5cm}ccccccccl}
\begin{deluxetable*}{@{\hspace{0cm}}l@{\hspace{-1cm}}c@{\hspace{.2cm}}c@{\hspace{0cm}}c@{\hspace{0cm}}c@{\hspace{0cm}}c@{\hspace{0cm}}c@{\hspace{0cm}}c@{\hspace{0cm}}c@{\hspace{0cm}}c@{\hspace{0cm}}c@{}c@{}c@{}}
\tabletypesize{\scriptsize}
%\vspace{-10cm}
%\tablecolumns{13} 
%\setlength{\columnsep}{1pt}
\tablewidth{\textwidth}
%\rotate{}
%\columnsep{1pt}
%\setlength{\tabcolsep}{2pt}
\tablecaption{Coma Berenices APOGEE Targets\label{restable}}
%\tablewidth{0pt}
\tablehead{
\colhead{\hspace{-.2cm}Index} & \colhead{\hspace{.3cm}2MASS ID} &  \colhead{H} & \colhead{RV\tablenotemark{a}} & \colhead{\hspace{-.3cm}$\sigma_{\mathrm{RV}}$} & \colhead{\hspace{-.2cm}T$_{\mathrm{eff}}$\tablenotemark{b}} & \colhead{\hspace{-.2cm}Kraus\tablenotemark{c}} & \colhead{\hspace{-.2cm}Merm. RV\tablenotemark{d}} & \colhead{\hspace{-.2cm}Spec\tablenotemark{c}} & \colhead{[Fe/H]\tablenotemark{e}} & \colhead{[Fe/H]\tablenotemark{e}} & \colhead{$v \sin i$} & \colhead{Notes} \\
&&& \colhead{\hspace{-.2cm}(km s$^{-1}$)} & \colhead{\hspace{-.3cm}(km s$^{-1}$)} & \colhead{\hspace{-.2cm}(K)} & \colhead{\hspace{-.2cm}Prob \%}  & \colhead{\hspace{-.2cm}(km s$^{-1}$)} & \colhead{\hspace{-.2cm}Type} & IRTF-K & IRTF-H & (km s$^{-1}$) &
}
\startdata
 0 & 12231356+2602185 & 11.52 & $ +2.84$ &   0.17 &  3390 &  98.10 & \ldots & M3.9 & $ -0.15$ & $ -0.06$ & $17.40\pm 3.30$ & New M Dwarf\\
 1 & 12255421+2651387 & 11.39 & $ +2.35$ &   0.28 &  3428 &  97.30 & \ldots & M2.7 & $ +0.01$ & $ -0.03$ & $ 4.40\pm 2.30$ & New M Dwarf\\
 2 & 12242665+2545077 & 10.18 & $ +2.54$ &   0.27 &  3433 &  88.50 & \ldots & M3.2 & $ +0.06$ & $ +0.03$ & $12.70\pm 2.50$ & New M Dwarf\\
 3 & 12264027+2718434 & 11.90 & $ +3.25$ &   0.20 &  3440 &  98.30 & \ldots & M2.8 & \ldots & \ldots & $ 9.10\pm 1.90$ & New M Dwarf\\ \smallskip 
 4 & 12201448+2526072 & 11.70 & $ +2.50$ &   0.18 &  3480 &  97.80 & \ldots & M2.5 & $ +0.01$ & $ -0.06$ & $ 8.10\pm 1.90$ & New M Dwarf\\
 5 & 12250262+2642382 & 11.03 & $ +1.20$ &   0.43 &  3485 &  98.80 & \ldots & M2.4 & $ -0.01$ & $ -0.05$ & $ 7.20\pm 1.80$ & New M Dwarf\\
 6 & 12214070+2707510 & 10.12 & $ -0.17$ &   0.34 &  3582 & \ldots & \ldots & \ldots & $ -0.11$ & $ -0.12$ & $14.00$\tablenotemark{g} & New K Dwarf;SB2\\
 7 & 12241121+2653166 & 10.27 & $ +0.70$ &   0.31 &  3699 & \ldots & \ldots & \ldots & $ -0.13$ & $ -0.06$ & $ 9.70\pm 1.20$ & New K Dwarf\\
 8 & 12265103+2616018 &  9.27 & $ +0.27$ &   0.13 &  4356 &  87.10 & $ -0.69$ & K4.8 & \ldots & \ldots & $ 8.00\pm 1.00$ & Arty 537 \\ \smallskip 
 9 & 12232820+2553400 &  9.35 & $ -0.07$ &   0.18 &  4370 &  95.90 & $ -0.20$ & K4.1 & \ldots & \ldots & $ 6.80\pm 1.00$ & Arty 278 \\
10 & 12211561+2609140 &  9.09 & $ +0.47$ &   0.15 &  4663 &  94.40 & $ -0.01$ & K3.5 & \ldots & \ldots & $ 7.50\pm 1.40$ & CJD 17 \\
11 & 12260547+2644385 &  7.69 & $ +6.23$ &   7.22 &  4864 &  93.20 & $ +1.03$ & K1.6 & \ldots & \ldots & $10.00$\tablenotemark{g} & Tr 120;SB2\\
12 & 12285643+2632573 &  8.77 & $ +1.00$ &   0.33 &  4932 &  92.30 & $ +0.83$ & K5.3 & \ldots & \ldots & $ 5.10\pm 1.00$ & Ta 20 \\
13 & 12223138+2549424 &  7.15 & $ +0.85$ &   1.15 &  5789\tablenotemark{f} & \ldots & $ -0.23$ & \ldots & \ldots & \ldots & $25.50\pm 5.20$ & Tr 92\\ \smallskip 
14 & 12230840+2551049 &  7.76 & $-21.35$ &  30.89 &  6022\tablenotemark{f} & 100.00 & $ -0.23$ & F9.7 & \ldots & \ldots & $13.50\pm 1.60$ & Tr 97;SB1\\
15 & 12234182+2636054 &  7.79 & $+12.30$ &   4.21 &  6156\tablenotemark{f} & 100.00 & $ +0.18$ & G5.5 & \ldots & \ldots & $<4.00$\tablenotemark{g} & Tr 102;SB2\\
16 & 12204557+2545572 &  7.74 & $ -0.47$ &   0.15 &  6225\tablenotemark{f} &  99.40 & $ -0.75$ & F7.8 & \ldots & \ldots & $ 4.90\pm 1.90$ & Tr 76 \\
17 & 12215616+2718342 &  7.40 & $ +0.82$ &   0.17 &  6351\tablenotemark{f} &  99.90 & $ -0.77$ & F4.2 & \ldots & \ldots & $16.00\pm 1.60$ & Tr 86 \\
18 & 12255195+2646359 &  7.30 & $ +0.83$ &   0.10 &  6650\tablenotemark{f} & 100.00 & $ +0.27$ & F2.5 & \ldots & \ldots & $14.00\pm 1.00$ & Tr 118 \\ \smallskip 
19 & 12234101+2658478 &  7.33 & $ +0.94$ &   0.21 &  6662\tablenotemark{f} & 100.00 & $ -0.17$ & F3.0 & \ldots & \ldots & $23.40\pm 1.70$ & Tr 101 
\enddata
\tablenotetext{a}{RV from SDSS-III DR10 public release.}
\tablenotetext{b}{T$_{\mathrm{eff}}$ estimated from empirical $V-K_s$ - T$_{\mathrm{eff}}$ relation from \citet{Boyajian:2012eu} with [Fe/H] = 0. The median absolute deviation for this fit reported by \citet{Boyajian:2012eu} is 49K, and we estimate the uncertainty in $V-K_s$ to be $\sim0.05$, corresponding to $\sim100-140$~K.}
\tablenotetext{c}{Membership probability and spectral type are based on SED fitting from \citet{Kraus:2007cx}. \citet{Kraus:2007cx} estimate the upper limit on the uncertainty in the spectral type to be three subclasses for stars from A0-G0 and 1 subclass for stars from G0-M6.}
\tablenotetext{d}{Radial velocity from \citet{Mermilliod:2008dx}.}
\tablenotetext{e}{The metallicity calibration uncertainty is $\pm 0.12$ dex.}
\tablenotetext{f}{$V-K_s$ is outside the well-calibrated range of \citet{Boyajian:2012eu} ($2<V-K_s<5$), so indicative temperatures from ASPCAP are shown. The estimated uncertainty for ASPCAP temperatures is $\sim150$~K, based on internal scatter \citep{Meszaros:2013va}.}
\tablenotetext{g}{The $v\sin i$ values for the spectroscopic binaries are estimated from template optimization during orbital fitting and are only indicative.}
\end{deluxetable*}
%\end{landscape}

\section{Target Characterization} 
In this section we describe further characterization of the Coma Ber targets, including rotational velocities derived from APOGEE spectra, secondary components of binary cluster members, metallicities and luminosity class verification using IRTF-SpeX medium-resolution spectra, and photometric variability from Kilodegree Extremely Little Telescope (KELT) monitoring.
\label{secchar}
\subsection{IRTF-SpeX: Metallicity and Luminosity Classes}
\begin{figure*}
\begin{center}
\includegraphics[scale=0.6]{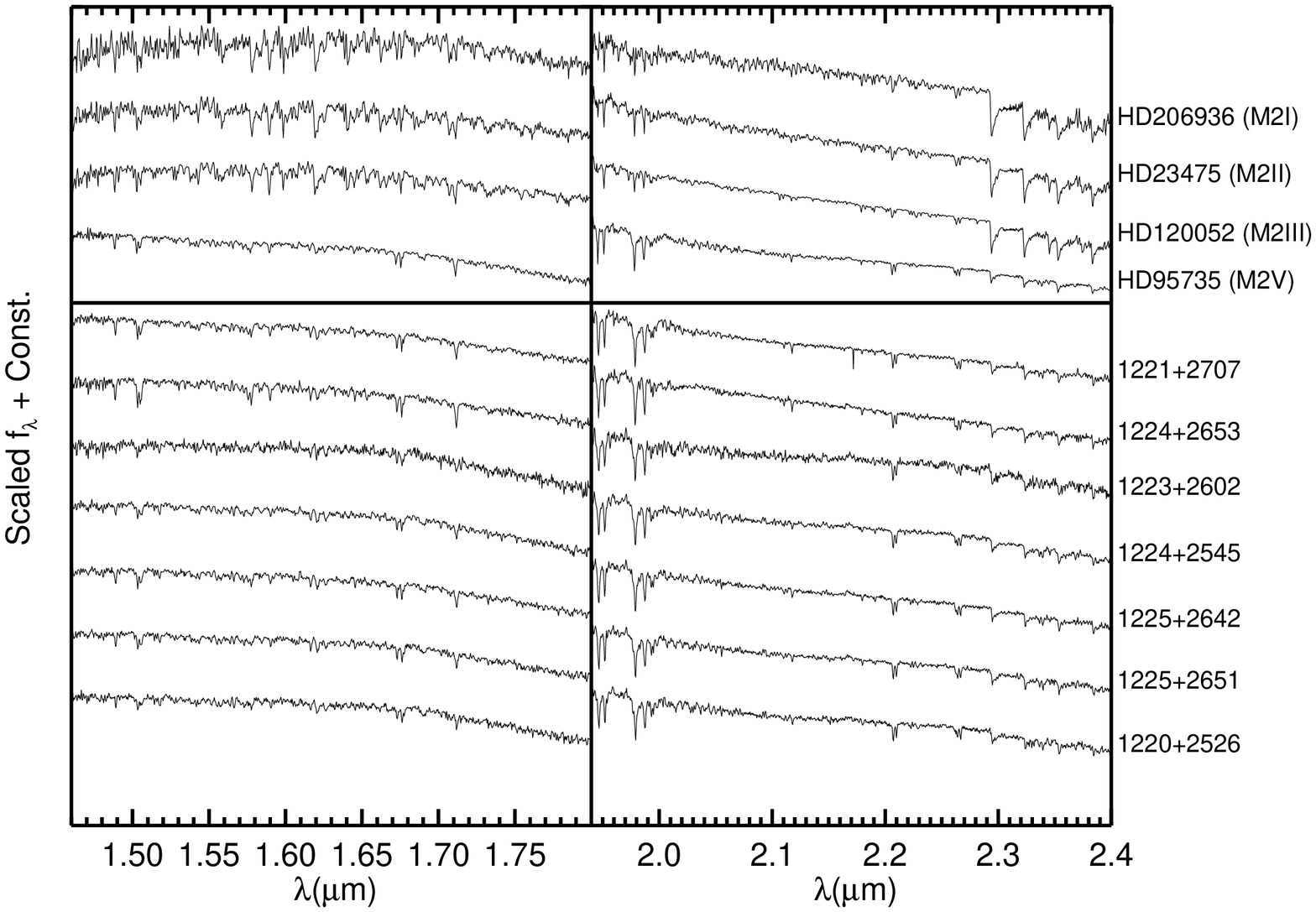} %fig 7
\caption{IRTF-SpeX \citep{2003PASP..115..362R} $R\sim 2000$ spectra of seven of the coolest APOGEE Coma Ber candidate targets (bottom). Comparison targets from the IRTF Cool Stars Library \citep{2009ApJS..185..289R} are shown at the top, as examples of different luminosity classes. The IDs displayed on the right are abbreviated 2MASS IDs, for full identification see Table~\ref{restable}}\label{figirtf}
\end{center}
\end{figure*}

On 2013 May 24--26, we observed seven of the eight new APOGEE low mass Coma Ber members with the IRTF-SpeX spectrograph \citep{2003PASP..115..362R}. These observations provide independent confirmation of the luminosity classes of these stars using extended spectral coverage and comparison with a library of stellar templates \citep{2009ApJS..185..289R}. We also measured metallicitoes for these targets as a part of our IRTF M dwarf metallicity determination program \citep{2012ApJ...747L..38T}. As described in \citet{2012ApJ...747L..38T}, we use an empirical metallicity calibration of a small number of features in the $H$ and $K$ bands to constrain metallicity for these stars. The metallicity measurements for the APOGEE Coma Ber targets, shown in Table \ref{restable}, indicate that these stars are of solar or slightly sub-solar metallicity. These measurements are not by themselves strong constraints on cluster membership, since most M dwarfs in the Solar neighborhood have similar metallicities. Nonetheless, the metallicity measurements for the new Coma Ber M dwarfs are in agreement with previous determinations of [Fe/H] $ = -0.05$ for Coma Ber \citep{CayreldeStrobel:1990vu,Friel:1992fc,Gratton:2000us}. The IRTF spectra also allow verifications of the luminosity class of the observed stars through comparison with the IRTF Spectral Library \citep{2009ApJS..185..289R}. These comparisons for the $H$ and $K$ bands, which contain the gravity-sensitive CO overtone bands, are shown in Figure \ref{figirtf}, and also independently confirm that these stars are indeed dwarfs.

\subsection{Rotational Velocities}
\label{secvsini}
We measured rotational velocities for all the Coma Ber targets using the method outlined in \citet{Deshpande:2013we}. Briefly, this technique involves maximizing the cross-correlation of a template spectrum with each epoch of a given target. The template is chosen from a grid of stellar models based on the outputs of the APOGEE reduction pipeline, which yield estimates of T$_{\mathrm{eff}}$, [Fe/H], and $\log(g)$. The set of models is rotationally broadened (with a range of kernels 3~km~s$^{-1}$ $\leq v\sin i \leq$~100~km~s$^{-1}$) and finally broadened according to the line spread function, which is also computed by the APOGEE pipeline. Each model is then cross-correlated against the observed spectrum, and the amplitude of the resulting correlation peak serves as the goodness of fit metric. Maximizing the correlation amplitude versus model $v\sin i$ gives the optimal rotation velocity for each visit spectrum. The output $v\sin i$ measurements are constrained to a precision of 2~km~s$^{-1}$ if there is only one visit, and are otherwise computed as the spread in measured $v\sin i$ across epochs (minimum 1~km~s$^{-1}$). In Table \ref{restable}, we report $v\sin i$ measurements with a floor of 4~km~s$^{-1}$. We do not use this method to measure $v\sin i$ for the double-lined spectroscopic binaries, so in Table \ref{restable} we instead report an estimated $v\sin i$ value for the primary star that maximizes the cross-correlation in our RV measurements.

\subsection{KELT Photometry}
The Coma Ber cluster was included in a field as part of the northern Kilodegree Extremely Little Telescope (KELT) survey, which is a wide-field photometric survey that can obtain $\leq$ 1$\%$ relative photometry for stars in the range of 8 $<$ V $<$ 11 mag and better than $\sim$10$\%$ for stars brighter than V$\sim$15 mag \citep{Pepper:2007ja,Pepper:2012ga}. The Coma Ber field has been observed by KELT-North for more than six years. The present data set comprises $\sim7000$~images from December 2006 to June 2012. Precise relative photometry was measured using a custom difference-imaging-analysis pipeline \citep[][R.~J. Siverd et al., in preparation]{Siverd:2012em,Pepper:2012ga}. All photometric data were processed in the normal KELT pipeline, this includes outlier data rejection and red noise trend filtering \citep{Kovacs:2005ea}.

\begin{figure}[h]
 \centering
 \includegraphics[scale=.5,angle=0]{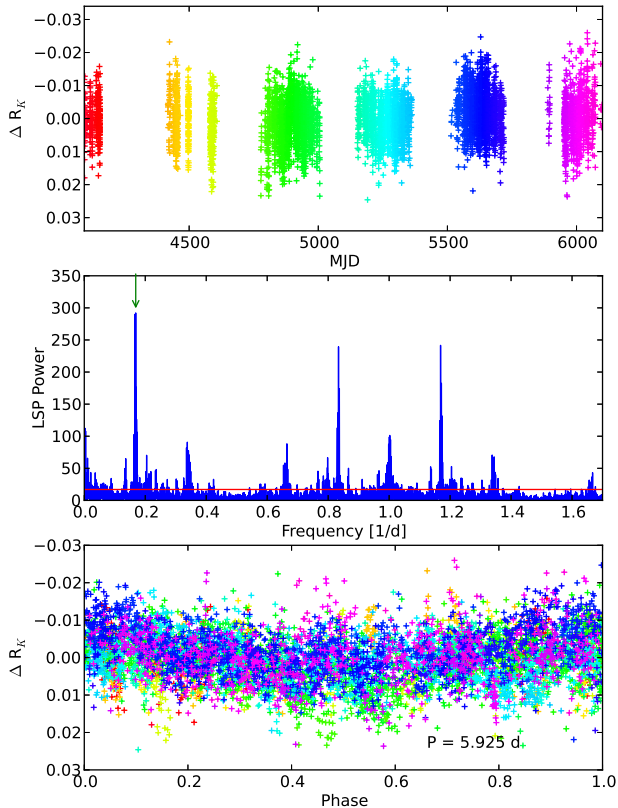}  %fig 8
 \caption{
   \label{fig.LCanalysis}  
   Example KELT light curve analysis for one Coma Ber cluster member. Top: Full KELT light curve with brightness of star plotted in relative KELT R$_{\rm K}$ instrumental magnitude. Color scale is relative to the date of observation given in MJD (JD$-$2400000~d). Middle: Lomb-Scargle periodogram for the KELT light curve. The power at a false alarm probability of 0.1\% is shown as a red horizontal line, and the tallest peak is indicated by the green arrow. Bottom: KELT light curve phased to the peak period in the periodogram (P$=$5.925~d). Color scale is the same as in the top panel.
   }
\end{figure}

In order to investigate photometric variability in Coma Ber, we matched the 385 APOGEE survey targets in the Coma Ber field with sources in the KELT database using a search radius of 23$\arcsec$ (i.e., one KELT pixel). For the resulting 80 matched objects, we identified and measured periodic variability in their light curves using a Lomb-Scargle Periodogram (LSP) \citep{Press:1989hb}. The significance of identified peaks in each LSP is tested by using a Monte Carlo simulation of random permutations of each light curve. We simulated 10000 random permutations of each KELT light curve, randomizing the photometric values while keeping the timestamps fixed. Applying the LSP to each simulated random light curve, we determine the height distribution for the tallest peaks in the LSP for these random light curves. We then compare the height of the star's LSP to this distribution to determine its false-alarm probability (FAP). We consider stars with significant periodic variability to have FAP $\leq$ 0.01, i.e., that each star's LSP has a peak with a height that is only seen per chance in less than 1\% of the 10000 randomized light curves. In Fig.~\ref{fig.LCanalysis}, we show examples of this analysis for a Coma Ber cluster member. We then visually inspected the KELT light curves and LSPs that passed this FAP cut. Objects with spurious low-frequency signals and related aliases were dropped. Errors on the measured periods were determined by calculating a 1$\sigma$ confidence level using the {\em post mortem} Schwarzenberg-Czerny method \citep{SchwarzenbergCzerny:1991wr}. Of the 80 total APOGEE targets in the Coma Ber field with KELT light curves, we find that 11 stars have significant photometric variability based on their LSPs (see Table~\ref{tab.Periods}). 

\begin{figure}[h]
 \centering
 \includegraphics[scale=.5,angle=0]{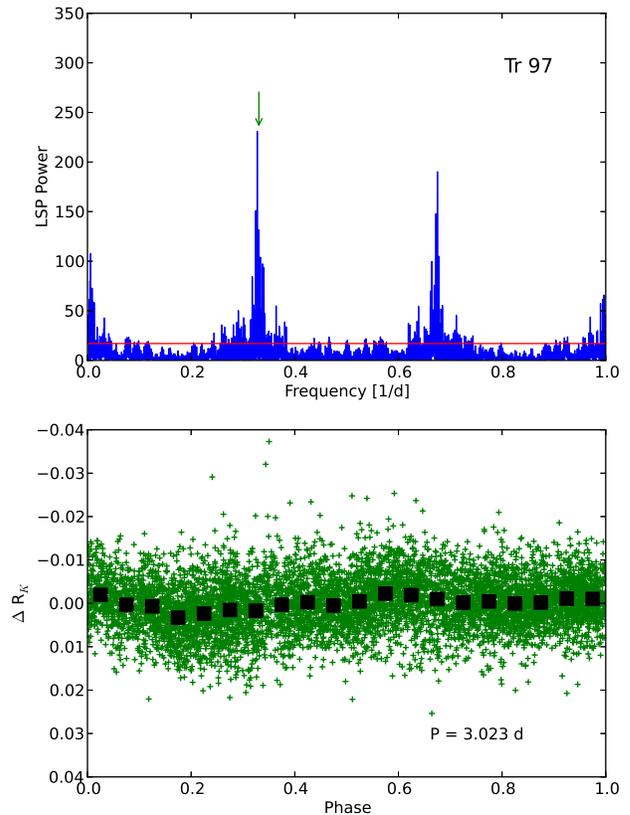} %fig 9
 \caption{
   \label{fig.LCbinary}  
   The periodogram (top) and phased KELT light curve (bottom) for Tr 98, the one spectroscopic binary in Coma Ber with unambiguous periodicity detected by KELT. The green arrow on the periodogram indicates the known orbital period of the system (3.02~d), and the power at a FAP$=$0.1\% is shown as a red horizontal line. The KELT light curve is phased to the orbital period, and over plotted in black symbols are the binned light curve values calculated using 20 equal size phase-bins.
   }
\end{figure}

\begin{deluxetable}{c c c c c}
\tabletypesize{\small}
%\tablecolumns{5}
\tablewidth{0pc}
\tablecaption{Measured Periodicity from KELT Light Curves \label{tab.Periods}}
\tablehead{
	\colhead{Index\tablenotemark{a}} & \colhead{2MASS} & \colhead{KELT} & \colhead{V} & \colhead{Period} \\
	\colhead{} & \colhead{ID}    & \colhead{ID}   & \colhead{} & \colhead{[d]}  
	}
\startdata
 8 & 12265103+2616018 & 03613 & $11.73$ & 11.78$\pm$0.0854  \\
  9 & 12232820+2553400 & 10817 & $10.51$ & 11.80$\pm$0.0273  \\
  10 & 12211561+2609140 & 02618 &$11.48$ & 1.098$\pm$0.0005 \\  
 11 & 12260547+2644385 & 01524 & $9.79$ & 8.810$\pm$0.0276 \\    
 12 & 12285643+2632573 & 04278 & $10.78$ & 9.483$\pm$0.0281  \\
 14 & 12230840+2551049 & 04561 & $9.09$ & 3.053$\pm$0.0017  \\
 16 & 12204557+2545572 & 03898 & $9.00$ & 5.925$\pm$0.0159 \\
 17 & 12215616+2718342 & 04083 & $8.50$ & 2.770$\pm$0.0016  \\
  \ldots & 12281397+2715328 & 05415 & $11.42$ & 5.595$\pm$0.0062 \\
 \ldots & 12245934+2540105 & 08031 & $10.21$ & 0.590$\pm$0.0001  \\
 \ldots & 12313367+2552132 & 12649 & $7.88$ & 0.854$\pm$0.0001
 \enddata
 \tablenotetext{a}{This index is defined in Table \ref{restable}.}
 \end{deluxetable}

Eight of these 11 variable sources are Coma Ber members. These stars extend from F to mid-K spectral types and have periods from $\sim$1--10 days. The variability seen in these cluster stars is thought to be due to photospheric heterogeneities, such as magnetically induced starspots, therefore, the periods determined for these stars are measurements of their rotation rates. The measurable photometric variability and rotation rates in these targets and the short rotation periods (P$<12$~d) suggests that they are relatively young, bolstering further the already high confidence of their membership in the young Coma Ber cluster.

 \citet{CollierCameron:2009ey} published an extensive study into the angular momentum distribution of the Coma Ber cluster using rotation periods determined from the SuperWASP survey. We found two of the cluster members with KELT rotation periods were also in this study. We find a similar period for 2MASS J12265103+2616018 in the KELT data (P$=$11.78~d) as \citeauthor{CollierCameron:2009ey} finds in the SuperWASP light curve (P$=$11.77~d), confirming the rotation period of this cluster member. And for 2MASS J12285643+2632573, we find a period (P$=$9.48~d) nearly twice that reported from the SuperWASP data for their 2004 observing seasons, but in agreement with their 2007 season data. \citeauthor{CollierCameron:2009ey} states that their 2004 period for this star was suspected to be half the true period as a result of the star having two dominant starspot groups on opposing stellar hemispheres. Our period for this star is measured with more data points and over a much longer time baseline ($\sim$7000 images over $\sim6$ years) and is thus less likely to be affected by short-timescale variation of the variability. This provides further evidence for the effect of period-halving due to symmetric starspot groups on this star.

In addition, we also closely inspected the KELT light curves and LSPs of the three identified binary cluster systems (Tr 97, Tr 102, and Tr 120) looking for photometric variability near the orbital period of the systems. Tr 97, a tight binary, shows significant periodicity at P$=$3.05~d (Figure~\ref{fig.LCbinary}), very near its orbital period \citep[3.02~d,][]{Mermilliod:2008dx}, indicative of a tidally synchronized orbital system. The two wider systems, Tr 102 and Tr 120, also show some level of periodicity in their light curves at their longer orbital periods (P$=$ 48.03078$\pm$0.00036~d and 294.785$\pm$0.031~d, respectively), but at lower significance. Tidal interactions are not predicted to be strong at these wider binary separations, and a search for eclipse-like variability did not return any conclusive results. Thus, we currently cannot definitively infer the source of this periodic variability.

%%%%%%%%%%%%%%%%%%%%%%%%%%%%%%%%%%%%%%%%%%%%

\subsection{Binary Cluster Members}
Three previously known binary members of Coma Ber were observed by APOGEE (Tr 97, Tr 102, and Tr 120), and one new binary member was discovered as well (2MASS 12214070+2707510). Each of these targets has three epochs of observation. The primary components of the three known spectroscopic binary systems have been monitored for many years with the CORAVEL spectrograph \citep{Mermilliod:2008dx}, but the secondaries have not previously been directly detected. We applied the TODCOR algorithm \citep{Zucker:1994fk} to attempt to detect these secondary components. The two-dimensional cross-correlation TODCOR algorithm simultaneously cross-correlates two stellar templates against the observed target blended spectrum and disentangles the stellar RVs of the two components with a maximum likelihood analysis \citep{Zucker:2003it}. TODCOR also derives a flux ratio for the system. Details concerning the implementation of the analysis can be found in \citet{2012ApJ...751L..31B}. In each case below, we use as stellar templates the solar metallicity BT Settl model spectra \citep{Allard:2012fp}  at the resolution and sampling of APOGEE. We successfully detected the secondary spectrum in Tr 102 and Tr 120, as well as the new binary 2MASS J12214070+2707510. The RVs measured from the APOGEE spectra for these target are listed in Table \ref{tabrv}. We note that during the first epoch of Tr 120, both components are near the systemic velocity. The two cross-correlation peaks cross-contaminate and fits to these peaks yield large RV uncertainties for both components. We did not detect the secondary component of Tr 97.

\begin{deluxetable}{lrrrr}
\tabletypesize{\small}
%\rotate{}
\tablecaption{Coma Berenices Spectroscopic Binary RV Measurements\label{tabrv}}
\tablewidth{0pt}
\tablehead{
\colhead{BJD} & \colhead{RV$_{\mathrm{A}}$}  & \colhead{$\sigma_{\mathrm{RV,A}}$}  & \colhead{RV$_{\mathrm{B}}$} & \colhead{$\sigma_{\mathrm{RV,B}}$} \\
& \colhead{(km s$^{-1}$)} & \colhead{(km s$^{-1}$)} & \colhead{(km s$^{-1}$)} & \colhead{(km s$^{-1}$)}
}
\startdata
\multicolumn{5}{c}{2MASS J12234182+2636054 (Tr 102)}\smallskip \\ 
2455939.956710  & $15.88$  &  $0.19$ & $-27.99$  &   $0.43$  \\
2455991.772350  & $14.97$ &   $0.19$ & $-25.72$  &   $0.46$  \\
2455999.796990   & $8.14$  &  $0.12$ & $-15.13$  &   $0.47$  \\ \hline
\multicolumn{5}{c}{2MASS J12260547+2644385 (Tr 120)}\smallskip \\
2455940.927860  & $-3.58$  &   $0.46$  &  $13.23$ &  $0.94$  \\
2455998.760300 &  $11.50$  &   $0.17$ &  $-17.79$  & $0.44$  \\
2456018.735160  & $10.92$  &   $0.17$  &  $-17.79$ &  $0.39$  \\ \hline
\multicolumn{5}{c}{2MASS J12214070+2707510}\smallskip \\
2455940.927860 &  $-1.47$  &   $0.23$  &  $2.25$  &   $0.25$  \\
2455998.760300 &  $12.44$ &    $0.19$ & $-12.79$  &   $0.14$  \\
2456018.735160  & $-9.40$  &   $0.22$  & $12.64$   &  $0.15$ 
\enddata
\end{deluxetable}

\begin{deluxetable*}{lrrrrrrr}
\tabletypesize{\small}
%\rotate{}
\tablecaption{Coma Berenices Spectroscopic Binary RV Fit Parameters\label{tabfit}}
\tablewidth{0pt}
\tablehead{\colhead{2MASS ID} & \colhead{T$_{\mathrm{eff,A}}$} & \colhead{T$_{\mathrm{eff,B}}$} & \colhead{$\log(g_{\mathrm{A}})$} & \colhead{$\log(g_{\mathrm{B}})$} & \colhead{$v\sin i_{\mathrm{A}}$} & \colhead{$F_{\mathrm{B}} / F_{\mathrm{A}}$} &  \colhead{Notes} \\
&\colhead{(K)} & \colhead{(K)} & \colhead{$\log(\mathrm{cm\ s}^{-2})$} & \colhead{$\log(\mathrm{cm\ s}^{-2})$} & \colhead{(km s$^{-1}$)} && }
\startdata
12234182+2636054 & $5500$ & $3800$ & $4.5$ & $5.0$ & $<4$ & $0.20$ & Tr 102 \\
12260547+2644385 & $4600$ & $3500$ & $4.5$ & $5.0$ & $10$ & $0.20$ & Tr 120 \\
12214070+2707510 & $3800$ & $3600$ & $4.5$ & $4.5$ & $14$ & $0.98$ & \ldots
\enddata
\end{deluxetable*}

Our TODCOR analysis \textbf{does} detect the secondary components of Tr 102 (2MASS J12234182+2636054, Target \#12), Tr 120 (2MASS J12260547+2644385, Target \# 9), and 2MASS J12214070+2707510 (Target \#6). In Table \ref{tabfit} we present the stellar template T$_{\mathrm{eff}}$, $\log(g)$, $v\sin i$, and flux ratio values that optimized the cross-correlation in the TODCOR analysis. For each system, broadening the secondary spectrum with a rotational velocity did not improve the cross-correlation, and for Tr 102 broadening the primary spectrum did not improve the cross-correlation either.

\begin{figure*}
\begin{center}
\includegraphics[scale=0.7]{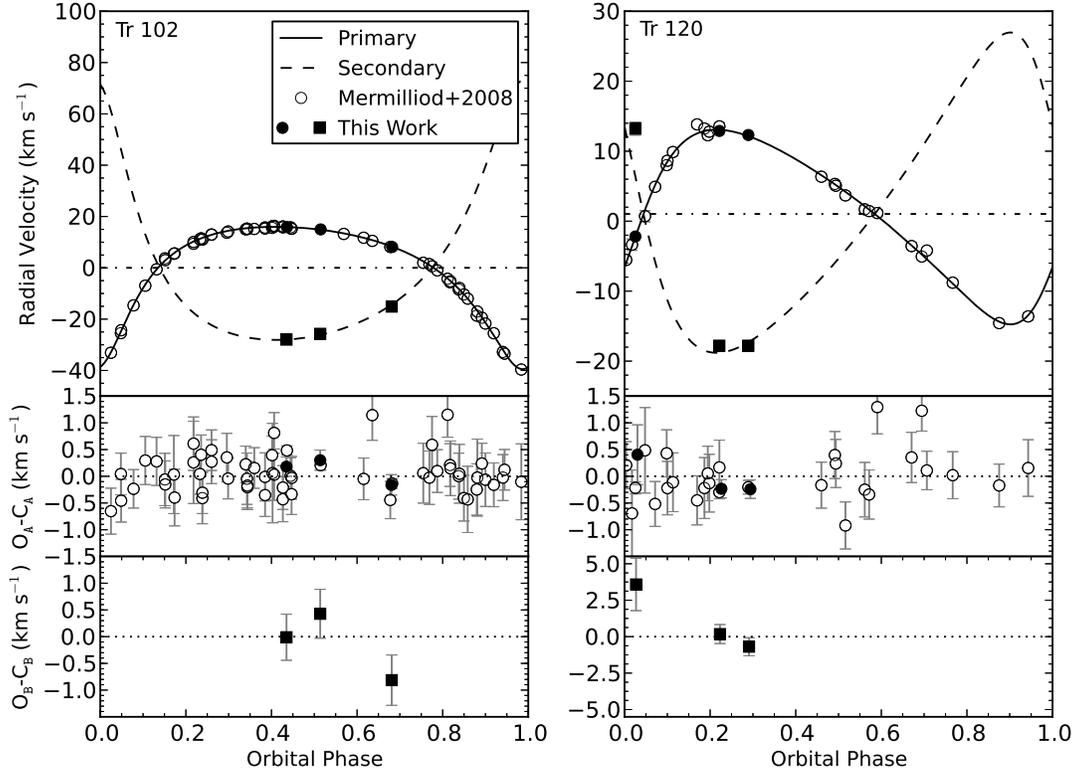} % fig 10
\caption{RV vs Orbital Phase for Tr 102 A \& B (left) and Tr 120 A \& B (right). Upper Panels: Filled circles and squares are the APOGEE derived V$_{\mathrm{A}}$ and V$_{\mathrm{B}}$, respectively. Open circles show V$_{\mathrm{A}}$ from \citet{Mermilliod:2008dx} that are offset to the APOGEE reference frame. Lower panels: Observed -- calculated residuals for the primaries and secondaries, respectively.}\label{figrv102}
\end{center}
\end{figure*}

We then used the RVLIN orbit fitting code \citep{Wright:2009jt} on the combined \citet{Mermilliod:2008dx} and APOGEE RV points to measure the orbital parameters, mass functions, and mass ratio for Tr 102 and Tr 120. These values are shown in Table \ref{taborb}, and the RV curves are shown in Figure \ref{figrv102}. These values are generally consistent with those presented in \citep{Mermilliod:2008dx}. For both systems we take advantage of the 30-year baseline covered by the \citep{Mermilliod:2008dx} and APOGEE datasets to provide updated periods and times of periastron passage. For the previously unknown binary system, 2MASS J12214070+2707510, we had only three APOGEE epochs, from which we derived a mass ratio and systemic ($\gamma$) velocity using the technique presented in \citet{Wilson:1941ev}. For this system, due to the small number of epochs, the determination of the systemic ($\gamma$) velocity and cluster membership confirmation were only possible because the system was resolved as a double-lined spectroscopic binary system.

We also note that the low-mass target 2MASS J12242665+2545077 (target ID 2) is more than a magnitude brighter than the isochrones in Figure \ref{figphot}, suggesting a possible binary. The APOGEE data for this target show marginal trends in both RV and $v\sin i$, however both are also consistent with the observational uncertainty. Further RV followup beyond the existing $\sim 80$~d baseline for this target could confirm or refute its possible binary nature.

\begin{deluxetable*}{lrrr}
\tabletypesize{\small}
\tablecaption{Coma Berenices Spectroscopic Binary Orbital Parameters for Objects Detected as SB2\label{taborb}}
\tablewidth{0pt}
\tablehead{\colhead{Parameter} & \colhead{12234182+2636054\tablenotemark{a}} & \colhead{12260547+2644385\tablenotemark{a}} & \colhead{12214070+2707510\tablenotemark{a}} \\
& \colhead{(Tr 102)} & \colhead{(Tr 120)} &
}
\startdata
$P$ (d) & $ 48.03058 \pm 0.00041$ & $294.78 \pm 0.03$ & \ldots \\
$T_0$ (HJD) & $2455774.964 \pm 0.08$ & $2455933.64 \pm 0.51 $ & \ldots \\
$\gamma$ (km s$^{-1}$) & $0.24 \pm 0.04$ & $1.13 \pm 0.15$ & $0.66 \pm 0.08$ \\
$e$ & $0.44987 \pm 0.00265$ & $0.323 \pm 0.016$ & \ldots \\
$\omega$ ($^{\circ}$) & $194.14 \pm 0.15$ & $245.50 \pm 0.74$ & \ldots \\
$K_1$ (km s$^{-1}$) & $27.75 \pm 0.12$ & $13.78 \pm 0.22$ & \ldots \\
$K_2$ (km s$^{-1}$) & $ 50.32 \pm 0.57$ & $22.90 \pm 0.73$ & \ldots \\
$M_{\mathrm{A}}\sin^{3}i$ ($M_{\odot}$) & $1.087 \pm 0.029$ & $0.798 \pm 0.060$ & \ldots \\
$M_{\mathrm{B}}\sin^{3}i$ ($M_{\odot}$) & $0.600 \pm 0.010$ & $0.480 \pm 0.025$ & \ldots \\
$M_{\mathrm{B}} / M_{\mathrm{A}}$ & $0.55 \pm 0.01$ & $0.60 \pm 0.02$ & $0.87 \pm 0.01$
\enddata
\tablenotetext{a}{2MASS ID}
\end{deluxetable*}

\section{Results}
\label{secres}

Our combined PM, RV, $\log (g)$, and photometric criteria yield 20 Coma Ber members, four of which are binary members with $\gamma$ velocities that confirm their cluster membership. The details of each target are shown in Table \ref{restable}. Of these 20 stars, 17 have been established as high probability cluster members through photometric and PM studies \citep{Kraus:2007cx,Casewell:2006ga}, and 12 have existing RV measurements that confirm their membership in Coma Ber \citep{Trumpler:1938vy,Mermilliod:2008dx}. We therefore confirm the membership of eight new stars in Coma Ber, including two previously not suspected to be Coma Ber members at all.

The six M dwarfs in this set are among those presented by \citet{Kraus:2007cx} as high probability candidates, based on an extensive PM and spectral energy distribution (SED) fitting analysis. Their APOGEE RV measurements fall within the $-3 < \mathrm{RV} < 4$~km~s$^{-1}$ limits we established. These are the first confirmed M dwarf members of Coma Ber. 

We also detect for the first time the low-mass secondary components of Tr 102 and Tr 120, previously known single-lined spectroscopic binaries in Coma Ber. For Tr 102, which has a primary spectral type of G0V \citep{Mermilliod:2008dx}, our measured mass ratio of $0.55\pm0.01$ indicates that the secondary component is likely a late K or early M dwarf. For Tr 120, the primary spectral type of K1V and a mass ratio of $0.60\pm0.02$ similarly indicates an M dwarf secondary component. The new spectroscopic binary 2MASS J12214070+2707510 appears to be another low mass system, with a mass ratio of $0.87\pm0.01$. These three double-lined spectroscopic binary systems more than double the number of such systems known in Coma Ber, after Tr 48 and Tr 150.

These new low-mass members of Coma Ber begin to populate the lower main-sequence for this cluster, and will provide important constraints for models of the cluster's dynamical history and mass function. Studies of other clusters have revealed the evolution of their low-mass populations \citep{Bouvier:2008kf,Wang:2011ki,Boudreault:2012fk}. Studies of the low-mass population of Coma Ber have suggested that it has not yet been stripped of its low-mass members \citep{Kraus:2007cx}. However these studies have relied on PM and photometry to identify low-mass members and have therefore suffered from contamination and incompleteness. A survey of the remaining candidate targets, possibly with APOGEE, would provide a well-constrained membership list with which to derive a cluster mass function. 

Moreover, as cluster members, the constrained age and composition of these new M dwarfs in Coma Ber will be useful for studies of low-mass stellar properties generally \citep[{e.g.}][]{Mamajek:2013tr}. If one of the Coma Ber systems with a low mass component is shown to be eclipsing, the stellar mass and radius measurements for this system would be a powerful test case for models of low-mass stellar radii. These models have difficulty reproducing the measured radii of these fully-convective low-mass stars \citep[e.g.][]{2012ApJ...757...42F,Boyajian:2012eu}. The known age and metallicity of the cluster are crucial for breaking degeneracies in the stellar models \citep[e.g.][]{Terrien:2012hj}.

Due to the limited number of targets and the small radius of the APOGEE field compared to the radius of Coma Ber, these low-mass stars likely represent only a fraction of the low-mass stars in Coma Ber. New studies, possibly with APOGEE, that can confirm or refute the membership of all candidate members within the entire area of Coma Ber could achieve the necessary completeness to gain leverage on models of the stellar population.

Due to the proximity and young age of Coma Ber, these M dwarf members may also be important targets for planet searches. Recent studies \citep{2013Natur.499...55M} have suggested that small planets can form and survive in clusters, and planets in the relatively young Coma Ber cluster \citep{Kharchenko:2005dw} may be in a stage of evolution corresponding to the Hadean Eon for Earth. The encouraging statistics of planets around M dwarfs from \textit{Kepler} \citep{Dressing:2013cq,Kopparapu:2013vu} suggests that the M dwarfs in Coma Ber could host planets in the Habitable Zone with larger RV and transit signatures than planets around FGK stars, potentially providing an important probe into planetary systems in the earliest stages of their evolution. \citet{Lloyd:2013wv} have outlined the advantages of a photometric survey designed to detect Jupiters around young stars, and the relative proximity of the Coma Ber open cluster puts our new K and M dwarf cluster members well within reach of planned photometric missions like \textit{TESS} \citep{Ricker:2010uo}.

\section{Conclusion}
Using APOGEE RVs, we confirm the discovery of the first low-mass members of the Coma Berenices cluster. This list includes the first six confirmed M dwarf members of Coma Berenices, which extends the list of confirmed members towards the bottom of the main sequence. We have used the APOGEE high-resolution $H$-band spectra to estimate $v\sin i$ values for these M dwarfs and the other members of Coma Berenices observed by APOGEE, and to detect for the first time the secondary components of the spectroscopic binaries Tr 102 and Tr 120. We also discover a new double lined spectroscopic binary system in Coma Berenices, 2MASS J12214070+2707510. We have obtained $R \sim 2000$ spectra from IRTF-SpeX that confirm the luminosity classes of seven of eight of the new low-mass Coma Ber members, and also indicate that their metallicities are approximately solar, in agreement with prior measurements of the cluster metallicity. Using KELT light curves, we find significant periodic photometric variability in eight higher-mass cluster members, which suggests their youth and allows us to measure rotational periods. This study demonstrates the power of the multiplexed APOGEE spectrograph for constraining cluster membership for late-type stars.

We thank the referee for insights and guidance that substantially improved this manuscript.

This work was partially supported by funding from the Center for Exoplanets and Habitable Worlds. The Center for Exoplanets and Habitable Worlds is supported by the Pennsylvania State University, the Eberly College of Science, and the Pennsylvania Space Grant Consortium. This work was also partially supported by the Penn State Astrobiology Research Center and the National Aeronautics and Space Administration (NASA) Astrobiology Institute. We acknowledge support from NSF grants AST 1006676, AST 1126413, and AST 1310885 in our pursuit of precision radial velocities in the NIR. PMF acknowledges support from NSF grant AST 1311835. This research has made use of the SIMBAD database, operated at CDS, Strasbourg, France. This publication makes use of data products from the Two Micron All Sky Survey, which is a joint project of the University of Massachusetts and the Infrared Processing and Analysis Center/California Institute of Technology, funded by the National Aeronautics and Space Administration and the National Science Foundation. 

This work was based on observations with the SDSS 2.5-meter telescope. Funding for SDSS-III has been provided by the Alfred P. Sloan Foundation, the Participating Institutions, the National Science Foundation, and the U.S. Department of Energy Office of Science. The SDSS-III web site is http://www.sdss3.org/. SDSS-III is managed by the Astrophysical Research Consortium for the Participating Institutions of the SDSS-III Collaboration including the University of Arizona, the Brazilian Participation Group, Brookhaven National Laboratory, University of Cambridge, Carnegie Mellon University, University of Florida, the French Participation Group, the German Participation Group, Harvard University, the Instituto de Astrofisica de Canarias, the Michigan State/Notre Dame/JINA Participation Group, Johns Hopkins University, Lawrence Berkeley National Laboratory, Max Planck Institute for Astrophysics, Max Planck Institute for Extraterrestrial Physics, New Mexico State University, New York University, Ohio State University, Pennsylvania State University, University of Portsmouth, Princeton University, the Spanish Participation Group, University of Tokyo, University of Utah, Vanderbilt University, University of Virginia, University of Washington, and Yale University.

This work was partially based on data from the Infrared Telescope Facility. The Infrared Telescope Facility is operated by the University of Hawaii under Cooperative Agreement no. NNX-08AE38A with the National Aeronautics and Space Administration, Science Mission Directorate, Planetary Astronomy Program.

%\bibliography{library}
\bibliographystyle{apj}
%\bibliography{library}

%\bibitem[{Frinchaboy {et al.} 2013, in prep}]{frinchaboy2013}
%Frinchaboy, P.~M. {et al.} 2013, in preparation

%\bibitem[{Gar\'{i}ca P\'{e}rez {et~al.} 2013, in prep}]{garcia2013}
%Gar\'{i}ca P\'{e}rez et al.\ 2013, in preparation
%
%\bibitem[{Nidever {et al.} 2013, in prep}]{nidever2013}
%Nidever, D.~L. {et al.} 2013, in preparation

%  
%\bibitem[{Shetrone, M., {et~al.} 2013, in prep}]{shetrone2013}
%Shetrone, M., et al.\ 2013, in preparation
%

%\bibitem[{Siverd, R., {et~al.} 2013, in prep}]{siverd2013}
%Siverd, R., {et~al.} 2013, in preparation

\end{document}